# Critical Currents, Pinning Forces and Irreversibility Fields in $(Y_xTm_{1-x})Ba_2Cu_3O_7$ Single Crystals with Columnar Defects in Fields up to 50 T

L. Trappeniers[1], J. Vanacken[1], L. Weckhuysen[1], K. Rosseel[1], A.Yu. Didyk[2], I.N. Goncharov[2], L.I. Leonyuk[3], W. Boon[1], F. Herlach[1], V.V. Moshchalkov[1] and Y. Bruynseraede[1]

*[1] Laboratorium voor Vaste-Stoffysica en Magnetisme, K.U.Leuven*
*Celestijnenlaan 200D, B-3001 Leuven, Belgium*

*[2] Laboratory of High Energy, Joint Institute for Nuclear Research, Dubna, Russia*

*[3] Faculty of Geology, M.S.U., Moscow 119899, Russia*

We have studied the influence of columnar defects, created by heavy-ion (Kr) irradiation with doses up to $6 \cdot 10^{11}$ Kr-ions/cm², on the superconducting critical parameters of single crystalline $(Y_xTm_{1-x})Ba_2Cu_3O_7$. Magnetisation measurements in pulsed fields up to 50 T in the temperature range 4.2 - 90 K revealed that: (i) in fields up to $m_0H \approx 20$ T the critical current $J_c(H,T)$ is considerably enhanced and (ii) down to temperatures $T \sim 40$ K the irreversibility field $H_{irr}(T)$ is strongly increased. The field range and magnitude of the $J_c(H,T)$ and $H_{irr}(T)$ enhancement increase with increasing irradiation dose. To interpret these observations, an effective matching field $B_{f*}$ was defined. Moreover, introducing columnar defects also changes the pinning force $f_p$ qualitatively. Due to stronger pinning of flux lines by the amorphous defects, the superconducting critical parameters largely exceed those associated with the defect structures in the unirradiated as-grown material: $J_{c,6 \cdot 10^{11} \ Kr-ions/cm²}(77 \ K, 5 \ T) \ ^3 \ 10 \times J_{c,ref}(77 \ K, 5 \ T)$.

**Corresponding Author:**

Lieven Trappeniers                          lieven.trappeniers@fys.kuleuven.ac.be
Laboratorium voor Vaste-Stoffysica en Magnetisme       tel. + (32) 16 32 71 98
Katholieke Universiteit Leuven              fax. + (32) 16 32 79 83
Celestijnenlaan 200D, B-3001 Heverlee, Belgium

## INTRODUCTION

The irreversible properties of superconductors, i.e. the critical current $j_c$ and the irreversibility field $H_{irr}$, can be improved by the inclusion of normal particles [1,2,3] or MgO nano-rods [4,5] during growth, by chemical substitution [6], twisters [7], by introducing arrays of sub micrometer holes [8,9,10] and by proton [11] and heavy-ion irradiation [12-18].

Superconducting cuprates irradiated with high-energy heavy ions form a particularly interesting system due to the presence of a distribution of amorphous columnar tracks. In these materials, correlated disorder associated with the tracks becomes dominant in flux line pinning and in vortex dynamics. The very efficient pinning of flux lines by the columnar defects is made possible by the similarity in their geometries; when the linear defects are nearly parallel to the applied magnetic field, vortices can be pinned over the entire length of the columnar defect. Models describing these systems [19] predict the appearance of an entangled flux liquid phase at high temperatures and a Bose-glass phase at lower temperatures and magnetic fields. At well-defined values of the applied magnetic field, an incompressible Mott-insulator phase shows up, characterised by very long relaxation times [19]. Up to now, the upper limit determined by the depairing critical current $j_c \sim 10^{13}$ A/m², is not yet reached and therefore the optimisation of the size and distribution of defects in high-temperature superconductors is a subject of continuing research.

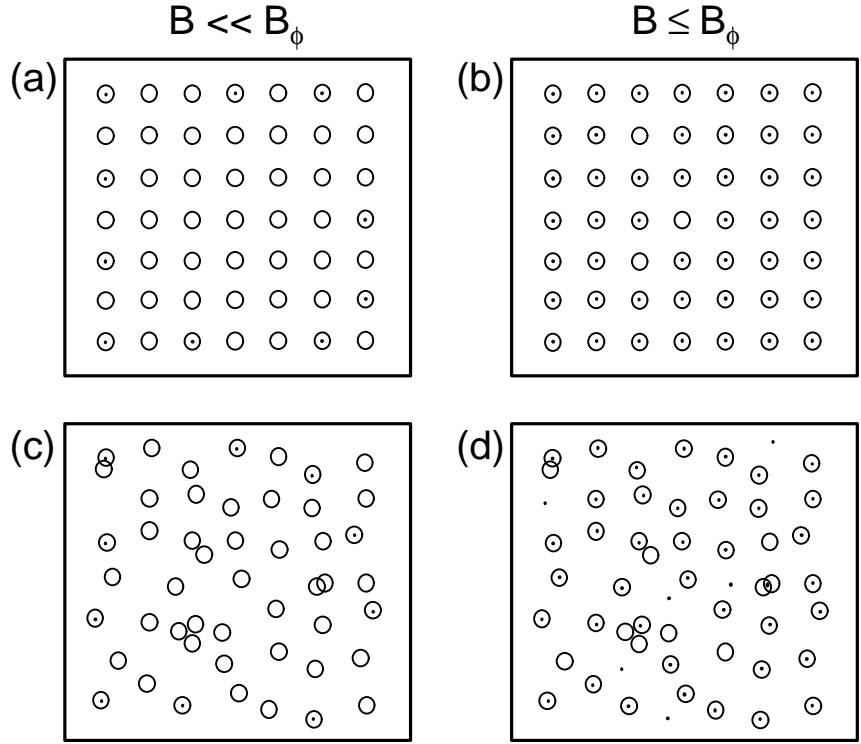

**Figure 1** *Schematic view of the pinning of vortices on a regular lattice of defects (a, b) and on a random distribution of defects (c, d) at $B << B_F (a, c)$ and $B £ B_F (b, d)$.*

For a *uniform distribution* of columnar defects almost all flux lines are pinned at low values of the magnetic field (see Fig. 1a). When the magnetic field is increased, the flux line density is equal to the averaged density of the defects at a matching field $B_F = n_f \cdot \Phi_0$ (with $n_f$ the density of the tracks) thus enabling optimal flux pinning (see Fig. 1b). The enhanced pinning at fields $B £ B_F$ is expected to improve the critical current and to induce a shift of the irreversibility line ($j_c(H_{irr}) \sim 0$) to higher magnetic fields. At fields $B > B_F$, additional flux lines are subjected to the repulsion from the potential barrier surrounding the vortices trapped by the columnar defects [20]. In the case of a *non-uniform distribution* of columnar defects, clusters of defects are present locally so that, due to vortex-vortex interactions, only a reduced number of the defects constituting the cluster is capable of pinning a flux line. This leads to an abundance of unpinned flux lines already at $B << B_F$ (see Fig. 1c,d).

Several parameters determine whether these vortices will be localised in the interstitial positions (between the flux lines strongly pinned by the columnar defects) or form multi-quanta vortices at the defect sites. According to the analysis of Buzdin et al. [21], the latter is possible when the radius $R$ of the columnar defects fulfils the condition:

$$R > \left( x_{ab} a_0^2 \right)^{1/3} \qquad (\lambda_{ab} >> d) \qquad (1a)$$

$$R > \left( x_{ab} l_{ab}^2 \right)^{1/3} \qquad (\lambda_{ab} £ d) \qquad (1b)$$

where $l_{ab}$ and $x_{ab}$ are respectively the in-plane penetration depth and coherence length, $a_0$ is the distance between the flux lines and $d$ is the average distance between the tracks. Equation (1b) corresponds to the case where the interaction between vortices can be neglected.

When the density of vortices is higher than the local density of columnar defects (and equation (1) is not fulfilled) the magnetic flux lines will go to those interstitial positions where the repulsion potential is minimal (see Fig. 1d). When the magnetic field is high enough to produce "quasi-bound" vortices at these interstitial positions, the relaxation of the magnetic response is dominated by the dynamics of interstitial vortices [22]. The efficiency of this interstitial pinning is temperature dependent, since it is related to the ratio between the inter-defect distance $d$ and the penetration depth $l_{ab}(T)$ (when $d/l_{ab} \ge 1$ the pinning potential becomes smoother). If the magnetic field is very high, the interaction between the flux lines becomes important and an Abrikosov lattice is created between the defects, resulting in a very small pinning potential [9,22].

In this paper we report on the influence of heavy ion irradiation on the critical current $J_c$, the pinning force $f_p$ and the irreversibility field $H_{irr}$ of $(Y_{0.14}Tm_{0.85})Ba_2Cu_3O_7$ single crystals at high magnetic fields. We verified to which extent the idealised picture of field matching on a uniform distribution of defects is still valid for a random distribution. It will be shown that in order to interpret the high-field data, an effective field $B_{f_\bullet}$ must be defined, corresponding to the density of defects *effectively available* for pinning.

## EXPERIMENTAL

Magnetisation measurements up to 50 T have been performed on a series of single crystals of $(Y_{0.14}Tm_{0.85})Ba_2Cu_3O_7$ with a $T_{c,mid} \sim 91$ K (Fig. 2). All crystals are from the same growth session, unirradiated reference samples as well as those irradiated with Kr. The single crystals have been grown by the self-flux method in $ZrO_2$:Y crucibles, starting from initial powders containing $Tm_2O_3$, $BaCO_3$ and $CuO$ in the ratio 3:25:72 [23] and have typical dimensions $1.2x1x0.023$ mm$^2$. The Tm/Y doping is a consequence of the $ZrO_2$:Y crucible used in the crystal growth [23] and has been determined by EDAX analysis. Subsequently, the crystals were irradiated at room temperature in the JINR U-400 cyclotron in Dubna with a 2.5 MeV/amu $^{84}$Kr-ion beam resulting in amorphous tracks. The irradiation doses varied between $0.75 \cdot 10^{11}$ Kr-ions/cm$^2$ and $6 \cdot 10^{11}$ Kr-ions/cm$^2$. The high-energy irradiation did not influence the $T_c$ values nor the transition width $DT_c$ as shown in Fig. 2.

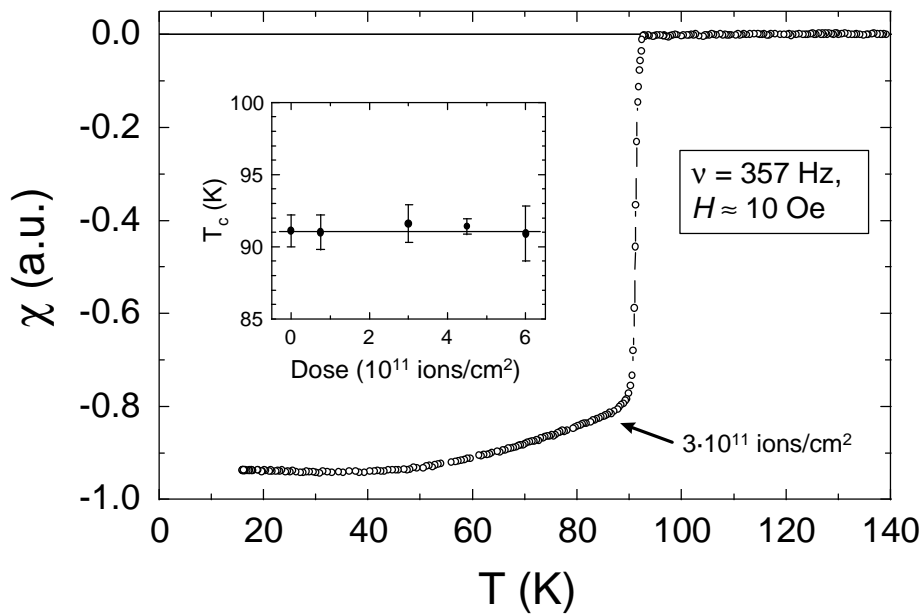

**Figure 2** *AC susceptibility vs. temperature for the $(Y_xTm_{1-x})Ba_2Cu_3O_7$ single crystal irradiated with $3 \cdot 10^{11}$ Kr ions/cm$^2$. The inset shows the $T_{c,mid}$ and $DT_c$ values (resp. $(T_c(90\%) + T_c(10\%))/2$ and $T_c(90\%) - T_c(10\%)$) as a function of the irradiation dose.*

The *morphology and the radius of the damage tracks* depends on the electronic stopping power $dE/dx$ which, by itself varies along the track. At high values of $dE/dx$, the tracks are continuous lines (region V in Fig. 12 of ref. [18]). When $dE/dx$ decreases, the tracks gradually change into semi-continuous lines (IV), a succession of cylinders (III) and spherical defects (II). In our case, using the TRIM code [24], it was calculated [25] that the track morphology starts as semi-continuous lines with a radius of 15 Å at the surface, changing to a more dashed structure deeper in the sample and transforming into spherical defects of radius 9 Å at about 10 µm. Depending on the estimate for the threshold value of $dE/dx$, the total penetration of the damage track is about 13 µm. The precise results of the TRIM calculation are summarised in Figure 3. From this discussion, it is clear that the notion *columnar track* must be used only when the track morphology is accounted for.

| $x$ | 0 μm (surface) | 6 μm | 8 μm | 10 μm |
|---|---|---|---|---|
| $f$ (nm) | 3.0 ± 0.5 | 2.7 ± 0.5 | 2.1 ± 0.4 | 1.8 ± 0.3 |
| $dE/dx$ (keV/nm) | 17.8 | 16.6 | 15.3 | 13.8 |
| morphology | 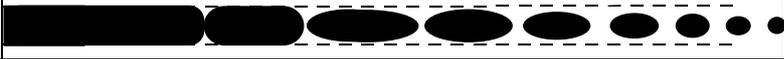 | | | |

**Figure 3** *Morphology of the damage track penetrating the sample from the surface. The track diameter Æ and the electronic stopping power dE/dx were calculated [25] using the TRIM-code [24]*

In our case, we have semi-continuous tracks, perforating the sample to at least 13 μm in depth (more than half of the sample) resulting in *blind holes*. The repercussions of such a partial penetration will be discussed.

The magnetisation measurements on these samples have been performed at the K.U.Leuven pulsed field facility [26] which allows to perform magnetisation measurements up to 60 tesla and at temperatures down to 35 mK. During a pulsed field magnetisation measurement, a 20 ms magnetic field pulse is applied parallel to the c-axis of the superconductor while at the same time, the susceptibility of the sample is measured with a calibrated susceptometer. The flat open inductive magnetisation sensor is calibrated [27] taking into account effects due to sample geometry, sample position on the sensor and the internal flux distribution inside the sample. The field dependence of the magnetic moment of the sample is then obtained by integrating the response of the detector. The sensitivity of this home-made susceptometer is better than $10^{-6}$ Am$^2$ at fields below 20 tesla and $10^{-5}$ Am$^2$ at higher fields [28]. Due to the short pulse duration of these high magnetic fields, huge sweep rates (more than 10 kT/s) are typical for such experiments. Moreover, since the duration of the field pulses is normally almost constant, the sweep rate is proportional to the peak field.

### RESULTS AND DISCUSSION

Since pulsed field magnetisation measurements of high temperature superconductors give more information than just the *M(H)* curve, it is appropriate to discuss the results of a typical experiment first. Figure 4 presents *M(H)* loops at 50 K (left) and 77 K (right) of the 3·10$^{11}$ Kr ion/cm$^2$ irradiated crystal measured during different field pulses. It is evident that the large sweep rates (see inset) in the rising-field branch of the *M(H)* loop cause a deformation of the hysteresis curve due to the dynamic response of the superconductor to strongly varying magnetic fields [29]. At *low temperatures* (Fig. 4a), the large shielding current induced in the crystal causes heating, resulting in a smaller magnetisation and a typical extra curvature in the *M(H)* loop at higher field pulses. At *higher temperatures* (Fig. 4b), there are no heating effects and the magnetic response is more sensitive to the intrinsic flux dynamics of the system. The sweep rate of the magnetic field during a magnetisation measurement can be used to calculate an effective voltage criterion which - by means of the current-voltage characteristic of the superconductor - determines the amplitude of the measured magnetic response [29,30]. These effects are less important in the decreasing field branch of the *M(H)* loop, where the sweep rate is lower and *M(H)* doesn't differ too much between the pulses of different amplitude. Taking into account these considerations, we have limited our interpretation of the experimental data to the branch of the hysteresis loop measured during decreasing magnetic field (upper branch of the magnetisation loop) and we reproduced the data with field pulses of different amplitude.

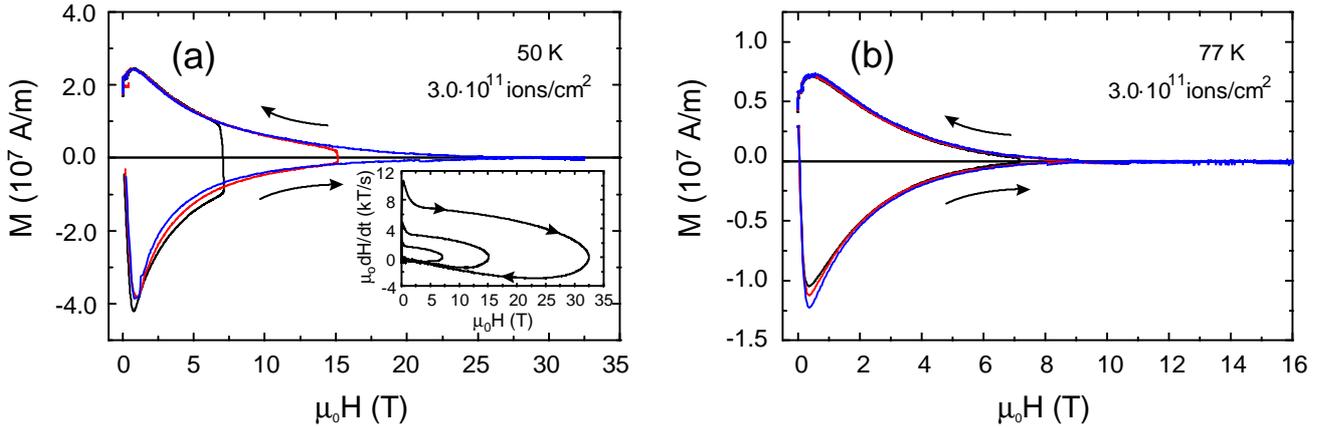

**Figure 4** Magnetisation of a $(Y_xTm_{1-x})Ba_2Cu_3O_7$ single crystal $(1.20x1.10\ mm^2)$ irradiated with $3\cdot10^{11}$ Kr ions/cm$^2$ measured in pulsed magnetic fields at 50 K (a) and 77 K (b). The inset of (a) shows the variation of the sweeprate of the magnetic field during the measurements shown.

Applying Bean's model [31] to the upper branch of the pulsed field magnetisation measurement gives the magnetic critical current density $j_{c,m} = \dfrac{3\left|M_+ - M_-\right|}{L} \approx \dfrac{6\left|M_-\right|}{L}$ where $L$ is the sample width and $M_+$ and $M_-$ are resp. the magnetisation at rising and lowering magnetic field. Figure 5 shows $j_{c,m}$ versus magnetic field for different irradiation doses, measured at 50 K and 77 K. It is clear that the presence of columnar tracks in the superconductor produces a change in the $j_{c,m}(H)$ behaviour; instead of a broad maximum, the curves show a pronounced peak at low fields (which is therefore truly related to the added amorphous tracks). The $j_c$ values substantially increase with higher doses and even at 5 tesla reach values of $1.2\cdot10^{11}$ A/m$^2$ at 50 K and $1.1\cdot10^{10}$ A/m$^2$ at 77 K. These values for $j_{c,m}$ are very high but they still lie well below the depairing critical current of $\sim 10^{13}$ A/m$^2$. At 50 K $j_{c,m}$ is enhanced, but the introduction of tracks does not induce a noticeable shift of the irreversibility field ($j_c(H_{irr}) \sim 0$). At a higher temperature, 77 K, the irreversibility field is shifted.

These observations agree with the expectation that, even if they are blind holes penetrating the sample more than halfway, the columnar tracks are such efficient pinning centres that their energy scale dominates the irreversible and dissipative response of the sample. This is because of their nature (normal regions) and their length along the flux line which enhances the scale of the pinning force and energy. Therefore, this system can be described with *a 2D model of pinning by strong pinning centres* (the tracks). From the moment a flux line is trapped by a columnar defect, it is pinned *strongly*, despite the weaker pinning by point defects in the bottom of the blind hole. The droplet-like region of the track (see Fig. 3) then is a natural transition to the pinning by point defects in the bottom of the blind hole.

The position and the width of the observed maximum in $j_c$ are clearly correlated with the dose of irradiation: for higher doses, the enhancement is stronger and an apparent shift of the maximum to higher fields occurs. An attempt can be made to understand this shift as the consequence of a certain commensurability between the mean distance between the defects (derived from the irradiation dose) and the inter flux line distance. Calculating the "conventional" (dose-equivalent) first matching field $B_F = n_f \cdot \Phi_0$ (with $n_f$ the averaged track density) at which each flux line of the vortex lattice ideally corresponds to one defect gives $B_F \sim 3.1$ T, 6.2 T and 12.4 T for the $1.5\cdot10^{11}$, the $3.0\cdot10^{11}$ and the $6.0\cdot10^{11}$ ions/cm$^2$ irradiated samples, respectively. This does *not* correspond to the experimental values for $B_{max}$ (the field at which the pinning force $f_p$ is extremal) of 2.0 T, 2.6 T and 3.0 T, respectively, but agrees well with earlier observations [14,32], where the ratio $B_{max}/B_F$ decreases to below 1 with increasing dose of irradiation. *This idealised picture of field matching on a uniform distribution of defects is thus not valid since, in reality, the defects are not distributed in an ordered way and locally, clusters of defects can be formed.* When the inter-defect distance in such a cluster is too small, not every defect can pin a flux line, due to the repulsive forces between the flux lines which are always tending to keep a regular order corresponding to a triangular flux line lattice. Consequently, *the number of effective pinning sites will be considerably lower than the one calculated from the irradiation dose.*

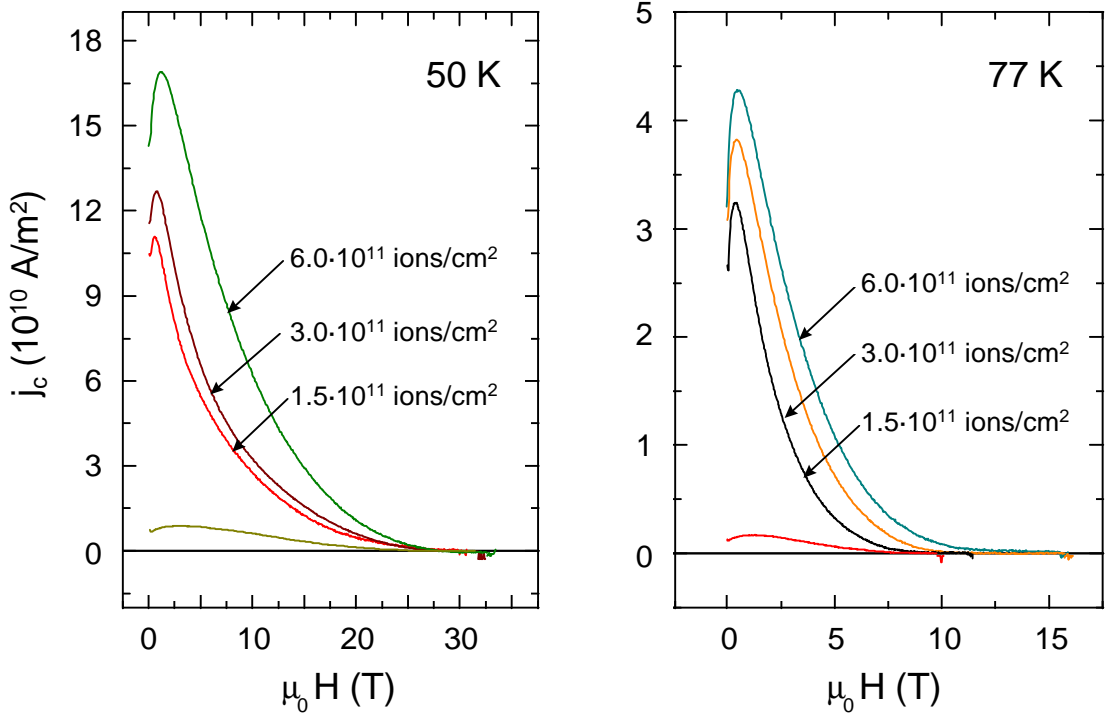

**Figure 5** *The critical current density at 50K (left) and 77K (right) for $Y_{0.14}Tm_{0.85}Ba_2Cu_3O_x$ crystals irradiated with different doses of Kr-ions. Defect densities of zero, $1.5 \cdot 10^{11}$, $3.0 \cdot 10^{11}$ and $6.0 \cdot 10^{11}$ Kr ions/cm$^2$ were used.*

In order to explain the unusual field dependence of the reversible magnetisation in irradiated high-$T_c$'s, Wahl et al [33] calculated the induction for the occupied columnar tracks by assuming a Poisson distribution of defects. In this work we will apply this concept to the *irreversible* magnetic properties of irradiated high-$T_c$ single crystals, reported here. A length scale $r = a_0\sqrt{x/(R+x)}$ is defined such that when two defects are at a distance smaller than $r$, only one of them can accommodate a flux line (see the inset in Fig. 6). Defining a grid with a spacing of $2^{\frac{1}{2}}r$ (see the inset) with an average number of $2r^2n_f$ defects per cell and assuming a Poisson distribution of defects leads to a probability $1 - P(0) = 1 - e^{-2r^2n_f} = 1 - e^{-ax}$ to find at least one defect in a cell, with $a = \dfrac{4}{\sqrt{3}}\dfrac{x}{R+x}$ and $x = \dfrac{B_f}{B}$. All defects in such a cell (1 or more) are capable of pinning 1 flux line *in total*; the others are not capable of pinning a flux line unless the length scale $r$ is changed (by changing the temperature or the applied field). This expression leads to a field dependent ratio between $B_{f\bullet}$ (corresponding to the concentration of defects *effectively available* for filling) and $B_F$ (ideal matching field corresponding to $n_f \cdot \Phi_0$) given in Eq. 2.

$$\frac{B_{f\bullet}}{B_f} = \frac{\left(1 - e^{-ax}\right)}{ax} \qquad (2)$$

This field dependent matching field $B_{f\bullet}$ is plotted in figure 6 for temperatures ranging from $t = T/T_c = 0$ to $t = 0.9$, taking $B_f = 6.2$ T and $R = 15$ Å; the bisecting $B_{applied} = B_{f\bullet}$ line corresponds to the filling of all *effectively available* defects. Above this line, the flux lines do not manage to fill all available defects (and added flux lines are trapped by the remaining available columnar defects). Below the bisecting line, all available defects are filled ($B_{applied} > B_{f\bullet}$) so that the additional flux lines are accommodated on interstitial positions since evaluating equation (1a) for our case leads to required fields above 350 tesla to generate multi-quanta vortices.

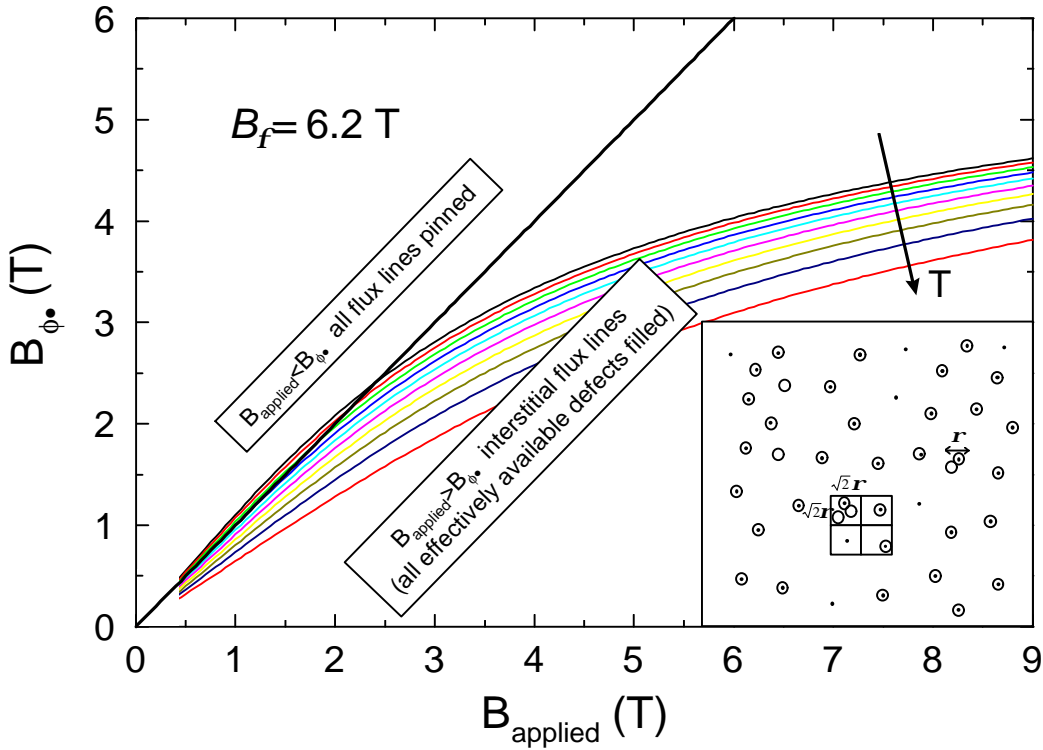

**Figure 6** *The field dependent matching field* $B_{f\bullet}$ *versus applied magnetic field. The averaged matching field* $B_f = n_f \cdot \Phi_0$ *was set to 6.2 T; the solid bisecting line corresponds to* $B_{applied} = B_{f\bullet}$. *The different curves are at temperatures between t = 0 (upper curve) and t = 0.9 (lower curve) in steps of 0.1. The inset is a schematic view of the pinning of flux lines on a random distribution of defects; defects at a distance closer than* **r** *can only pin a single vortex line per cluster.*

As introduced above, the effective matching field $B_{f\bullet}$ is always smaller than the dose-equivalent field $B_f$, the number of effectively available defects always being smaller than the total number of damage tracks. Sweeping up the magnetic field unleashes a process of continuous creation of new pinning sites (by changing the length scale **r**) which is accompanied by a continuous filling of these sites with the incoming flux lines. The slope of the $B_{f\bullet}(B_{applied})$ curve then determines the fraction of added flux lines that is pinned by the columnar tracks. If this fraction equals 1 (bisecting line), all flux lines are pinned. If the curves saturate (slope << 1), an increasing number of flux lines will be positioned at the interstitial sites. The irreversible response of the sample is thus not directly determined by the value of $B_f$ but rather by the slope of the $B_{f\bullet}(B_{applied})$ curve at $B_{applied}$, where $B_{applied}$ can be much higher than $B_f$.

It is clear that in order to enhance the irreversible properties it is favourable to operate in a regime where the superconductor is close to the bisecting $B_{applied} = B_{f\bullet}$ line. Once the applied field exceeds $B_{f\bullet}$, interstitial flux lines, which are more mobile [8,9], give rise to a monotonously decreasing $j_{c,m}$. This was recently observed [8] and also confirmed by flux-gradient driven molecular dynamics simulations of interacting vortices on a random distribution of defects [34].

Figure 7 represents a situation which is relevant for our experiments. In this plot, $B_{f\bullet}$ is given for different average $B_f$ values (derived from the experimental irradiation doses) for $T = 50$ K (left) and $T = 77$ K (right). We will use these plots to qualitatively explain the features of the $j_{c,m}(H)$ plots in figure 5. It is clear that, in our experimental data, we are always in the regime $B_{applied} > B_{f\bullet}$ (all available defects filled) such that interstitial flux lines are present. At 77 K, even at low fields a fraction of the incoming flux lines is occupying interstitial positions even though the filling of the available defects continues (the $B_{f\bullet}$ vs. $B_{applied}$ curves are not saturated). At the lower temperature of 50 K ($t = 0.54$), the superconductor is close to the special case where the initial slope of the $B_{f\bullet}(B_{applied})$ curve coincides with the $B_{applied} = B_{f\bullet}$ line. Thus, for the $6 \cdot 10^{11}$ Kr ion/cm² sample with a $B_F = 12.4$ tesla we are in the desirable $B_{applied} = B_{f\bullet}$ situation for fields up to ~ 4 tesla, whereas for the samples with a lower $B_F$, the $B_{f\bullet}(B_{applied})$ curve deviates from this optimal case. The fact that at 50 K the $6 \cdot 10^{11}$ Kr ion/cm² sample is in the $B_{applied} = B_{f\bullet}$ regime for such a large field region is reflected in the $j_{c,m}(H)$ plots in figure 5; the enhancement of $j_{c,m}(H)$ by increasing the irradiation dose is clearly much stronger at 50 K (left).

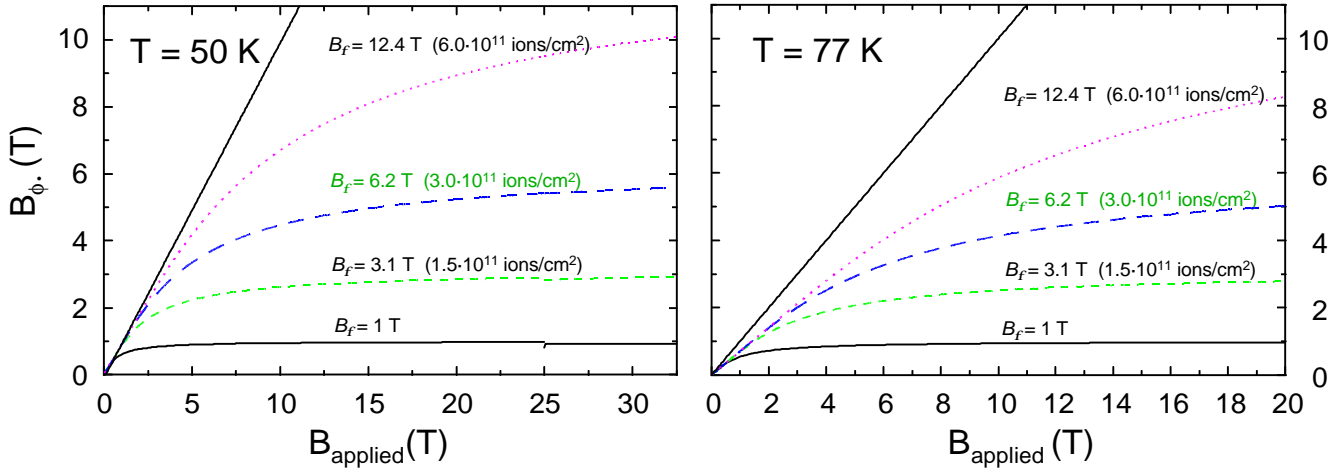

**Figure 7** *The field dependent matching field $B_{f\bullet}$, plotted versus applied magnetic field at 50 Kelvin and 77 Kelvin. The curves are for different averaged matching fields $B_f$ of 1 T, 3.1 T, 6.2 T and 12.4 T respectively. The solid bisecting line corresponds to $B_{applied} = B_{f\bullet}$.*

Not only the change in $j_{c,m}$ by adding columnar defects can be accounted for, also the observed shift of the irreversibility line can be explained by turning to an effective matching field $B_{f\bullet}$ which itself is dependent upon the applied magnetic field.

At 50 K (Fig. 5, left) we observe that the introduction of columnar tracks results in an increase of $j_{c,m}$ , *not* accompanied by a large increase of $H_{irr}$ . At 77 K however, the irreversibility field is increased *significantly* (Fig. 5, right) from 6.6 T (unirradiated) to 10.8 T (at the highest dose). This behaviour can be explained by looking at $B_{f\bullet}(B_{applied})$ in figure 7.

At 50 K, around $H_{irr}$ the $B_{f\bullet}(B_{applied})$ curve is saturating which implies that the incoming flux lines are located mainly at interstitial positions; the irreversible properties in this region thus reflect the intrinsic pinning properties of the unirradiated reference crystal and the observed $H_{irr}$ is the irreversibility field of the unirradiated reference sample. At higher fields - and hence at lower temperatures - a still increasing number of flux lines cannot be pinned by the columnar defects and these flux lines find themselves at the interstices. Therefore the critical current decreases (Fig. 5, left) and the irreversibility line of the unirradiated sample is completely recovered. In this region, only a very high density of amorphous defects will result in shifting the irreversibility line, as observed experimentally.

At 77 K (Fig. 7, right) we are in the regime where the $B_{f\bullet}(B_{applied})$ curves have a finite slope and a finite fraction of the flux lines is continuously occupying the increasing number of effectively available defects while only the other part is placed at the interstitial positions; the irreversibility line is therefore pronouncedly dependent on the irradiation dose, as observed experimentally.

The overall behaviour of the irreversibility line is summarised in figure 8. Within a well-defined area of the *H-T* plane, the experimental irreversibility line (defined by the resolution of the experiment) in this figure shows that the enhancement of $j_{c,m}$ due to the presence of the linear amorphous tracks is accompanied by a strong increase of $H_{irr}$ (see inset at 77 K). This agrees with our expectation that in the vicinity of the averaged $B_f$ flux pinning is more efficient. However, from figure 8, it is clear that *this enhancement extends far above the dose-equivalent field $B_f$*, another argument in favour of the refined interpretation in terms of $B_{f\bullet}$. In the low-field region of the *H-T* plane, where temperatures are rather high, the increasing coherence length results in pinning by the tracks (25 Å to 35 Å) to become less efficient and again the irreversibility line shifts back to the irreversibility line of the unirradiated crystal.

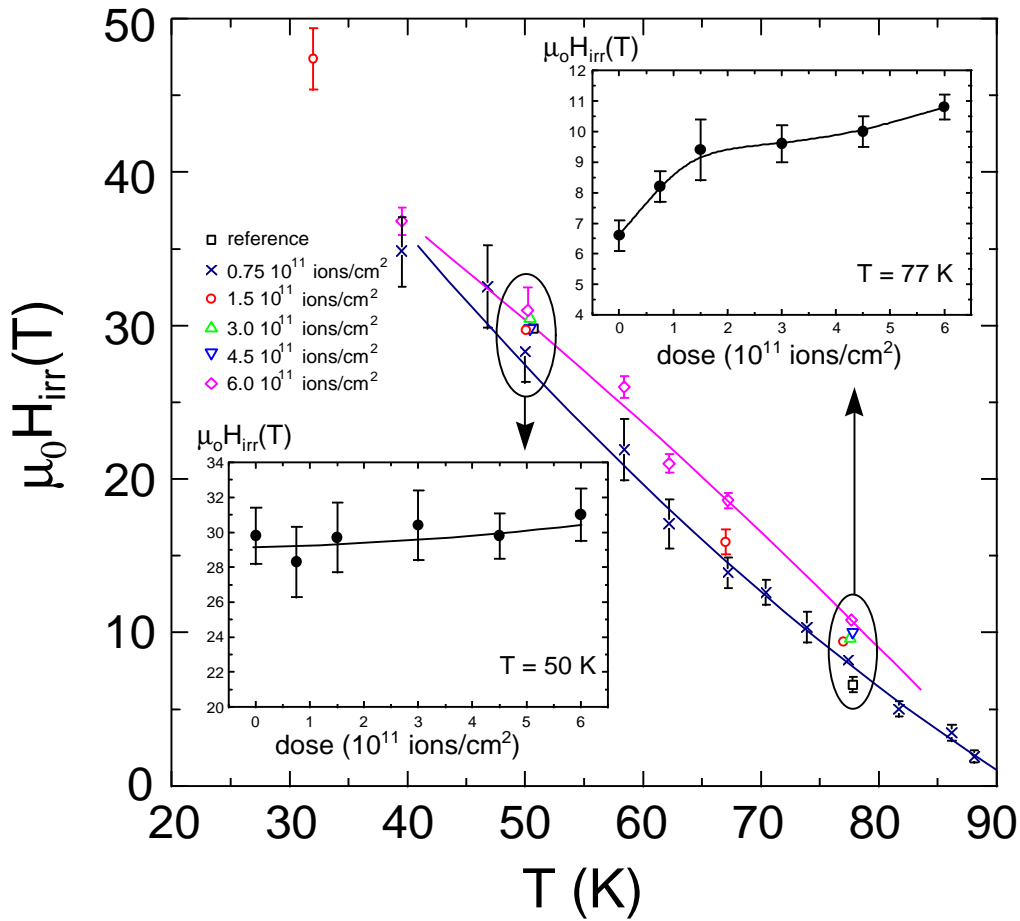

***Figure 8*** *The irreversibility field $H_{irr}(T)$ for the whole set of $Y_{0.14}Tm_{0.85}Ba_2Cu_3O_x$ crystals; the irradiation varies from 0 to $6 \cdot 10^{11}$ ions/cm$^2$. The insets represent $H_{irr}$ versus irradiation dose at fixed temperature.*

Additional information about the influence of linear amorphous defects can be obtained by looking at the pinning force associated with them. Figure 9 presents the pinning force $f_p(H)$ at 50 K and 77 K for all the samples. These graphs were obtained from the $j_{c,m}(H)$ plots in figure 5 by applying $f_h(B) = j_c(B) \cdot B$. Introducing linear amorphous defects clearly increases the pinning forces acting on the flux lines but also changes the field dependence qualitatively. This is clear when the pinning forces are normalised with respect to $B_{max}$ and $f_{p,max}$ (lower plots of figure 9). The normalised pinning force for the unirradiated sample clearly deviates from those for the irradiated crystals. At low magnetic fields, the pinning force in the irradiated samples follows one universal curve indicating that a unique pinning mechanism is applicable in that area. At higher fields, this scaling is poor. The data for the reduced pinning force have been compared with the various functional relationships proposed in the literature [35,36] assuming direct summation of the pinning forces; no agreement was found. This could either be due to the increased interaction between the flux lines at high fields or to the mixture of pinning at interstitial positions and defects sites. It should be mentioned that the $f_h(B) = j_c(B) \cdot B$ procedure magnifies experimental errors and that a minor deformation of the $f_p(H)$ curve can occur as a consequence of the sweep rate effects discussed above.

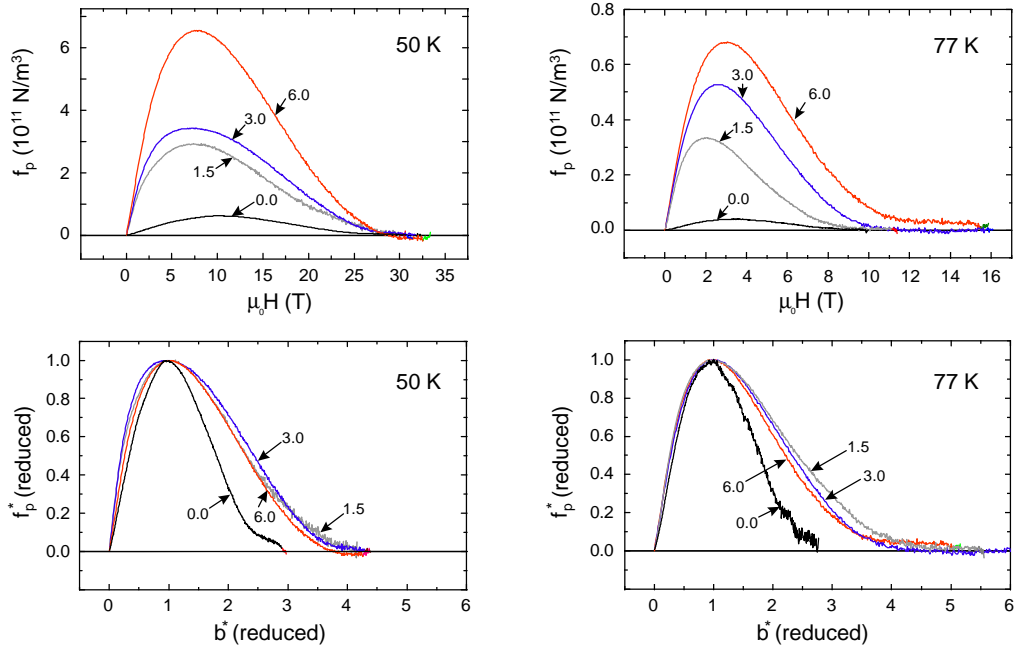

**Figure 9** *The pinning forces for all the samples at 50 K (left) and 77 K (right) on an absolute scale (upper plots) and a reduced scale (lower plots). The pinning forces were obtained from the critical current density.*

## CONCLUSIONS

The irradiated $(Y_{0.14}Tm_{0.85})Ba_2Cu_3O_7$ single crystals contain *random distributions* of both correlated (columnar defects) and uncorrelated pinning centres (Y/Tm point defects, oxygen deficiency or local stress). Our high-field study of these samples made it possible to determine the magnitude of the expected enhancement of the critical current and the irreversibility line at low temperatures and to study the origin of the microscopic pinning force over a large field range.

It has been shown that, even at fields above the averaged matching field $B_F$, the presence of linear amorphous tracks in $(Y_{0.14}Tm_{0.85})Ba_2Cu_3O_7$ single crystals results in a distinct enhancement of the critical current, well above the $j_c$ value of the unirradiated single crystal. This was interpreted in terms of an effective matching field $B_{f_\bullet}$, corresponding to the density of columnar defects *effectively available* for pinning, which by itself is dependent upon the applied magnetic field.

At fields above $B_F$ the irreversibility line shifts to higher temperatures. The magnitude and the position in the *H-T* diagram of the enhancement of the superconducting critical parameters $j_c$ and $H_{irr}$ depend on the dose of the heavy ion irradiation.

At fairly high doses of $6.0 \cdot 10^{11}$ ions/cm$^2$ the pinning properties of the material are improved and the limit where bulk superconductivity itself is suppressed has not yet been reached.

This shows that the influence of amorphous tracks on the irreversible properties of these $(Y_{0.14}Tm_{0.85})Ba_2Cu_3O_7$ single crystals is limited to a distinct region of the H-T phase diagram and that even at magnetic fields above $B_F$ an enhancement of the critical current and the irreversibility field can be observed. At higher magnetic fields the critical current returns to the value of the reference samples and the irreversibility line coincides with the depinning line for the unirradiated crystal.

## ACKNOWLEDGEMENTS

This work is supported by the FWO-Vlaanderen, the Flemish GOA and the Belgian IUAP programs. The co-operation with JINR and MSU was possible thanks to the European INTAS-94-3562 project. J.V., L.W. and K.R. are supported by the FWO-Vlaanderen. L.T. is a Research Fellow supported by the Flemish Institute for the Stimulation of Scientific and Technological Research in Industry (IWT).